\documentclass[11pt,bibnotes,aps,prl,twocolumn,reprint]{revtex4-1}
\usepackage{graphicx}
\usepackage{amsmath}

\begin{document}
\title{Negative dynamic Drude conductivity in pumped graphene}
\author{D. Svintsov}
\affiliation
{Institute of Physics and Technology, Russian Academy of Science, Moscow 117218}
\affiliation
{Moscow Institute of Physics and Technology, Dolgoprudny 141700, Russia}

\author{V. Ryzhii}
\affiliation 
{Research Institute for Electrical Communication, Tohoku University, Sendai 980-8577, Japan}
\affiliation {Center for Photonics and Infrared Engineering,
Bauman Moscow State Technical University, Moscow 105005, Russia}

\author{T. Otsuji}
\affiliation
{Research Institute for Electrical Communication, Tohoku University, Sendai 980-8577,  Japan} 

\begin{abstract}
We theoretically reveal a new mechanism of light amplification in graphene under the conditions of interband population inversion. It is enabled by the indirect interband transitions, with the photon emission preceded or followed by the scattering on disorder. The emerging contribution to the optical conductivity, which we call the interband Drude conductivity, appears to be negative for the photon energies below the double quasi-Fermi energy of pumped electrons and holes. We find that for the Gaussian correlated distribution of scattering centers, the real part of the {\it net} Drude conductivity (interband plus intraband) can be negative in the terahertz and near-infrared frequency ranges, while the radiation amplification by a single graphene sheet can exceed 2.3\%. 

\end{abstract}

\maketitle
Graphene under the conditions of interband pumping has attracted considerable interest~\cite{Malic-Book} due to the rich physics including nonlinear photoresponse~\cite{High-electric-field-dynamics}, collinear relaxation and recombination~\cite{Collinear}, anomalous carrier diffusion~\cite{Anomalous-diffusion}, and self-excitation of surface plasmons~\cite{Gain-enhancement}. The emergence of population inversion~\cite{Satou-Vasko-Ryzhii-pop-inv,Nature-pop-inv,Malic-pop-inv} and negative interband dynamic conductivity in pumped graphene enables the amplification of radiation, particularly, in the terahertz (THz) range~\cite{Ryzhii-NDC}. The experimental observation of coherent radiation amplification~\cite{THz-amplification-experiment,PRL-experiment} supports an idea of graphene-based THz-laser~\cite{Feasibility-of-THZ-lasing,Ryzhii-injection-laser}.
Its full-scale realization faces, however, a number of challenges. First, the coefficient of the interband amplification by a clean layer of pumped graphene cannot exceed 2.3\%~\cite{Ryzhii-NDC}, which is inseparably linked with the universal optical conductivity of graphene~\cite{Measurement-of-OC}. Second, the radiation amplification associated with the direct interband electron transitions competes with the intraband Drude absorption~\cite{Threshold-of-pop-inv, Loss-compensated-materials} which is inversely proportional to the frequency squared. 
The intraband photon absorption can be assisted by the processes of electron-phonon~\cite{Photoconductivity-graphene}, electron-impurity, and carrier-carrier~\cite{CCScattering,CCScattering2} scattering.

\begin{figure}[ht]
\center{\includegraphics[width=0.95\linewidth]{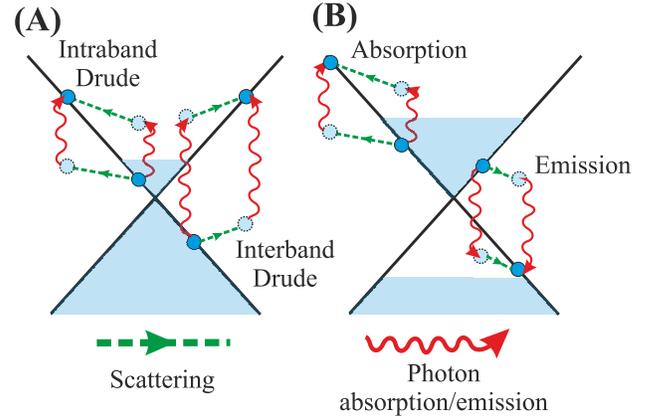}}
\caption{Schematic view of indirect electron transitions in graphene at equilibrium (A) and under conditions of population inversion (B). A common indirect transition involves electron scattering and photon absorption occurring within one band. An indirect interband transition involves {\it interband} scattering followed or preceded by photon absorption or emission.}
\label{F1}
\end{figure}

Considering the optical conductivity of direct-gap semiconductors, one typically accounts only for direct interband and indirect intraband electron transitions (the latter shown in Fig.~\ref{F1}A). There also exist indirect interband transitions, with the photon absorption (or emission) followed (or preceded) by the disorder (impurity or phonon) scattering (Fig.~\ref{F1}, A and B). In the direct gap semiconductors these are, however, generally excluded from concern, as they are the next-order processes compared to the direct interband transitions. In semiconductors possessing band gap $E_G$, indirect interband transitions are also less probable than indirect intraband ones. They can appear only at frequencies $\omega > E_G/\hbar$ wherein the Drude factor $1/\omega^2$ is small. This is not the case of gapless graphene, where it is reasonable to consider indirect intra- and interband transitions on equal footing. 

In the present letter we show that certain sources of scattering (namely, Gaussian correlated disorder) can result in dynamic interband Drude conductivity even higher than intraband one. Under conditions of population inversion, the interband scattering-assisted photon emission complements to the direct interband emission. The radiation amplification coefficient by such a graphene layer can exceed the 'clean' limit of 2.3\%.

Consider electrons in a graphene layer interacting with in-plane electromagnetic field described by vector-potential ${\bf A} = {\bf A}_0 \cos \omega t$ and random scattering potential $V_S({\bf r})$. The Hamiltonian of the system reads
\begin{gather}
\label{Hamiltonian}
\hat H = \hat H_0 + \hat V_F + \hat V_S ({\bf r}),\\
\hat H_0 = v_0 ({\boldsymbol \sigma},\hat{\bf p}),\\
\label{Field}
\hat V_F = -\frac{e v_0}{c} {\boldsymbol \sigma}{\bf A}_0 \cos\omega t,\\
\label{Scattering}
\hat V_S({\bf r}) = \sum\limits_{{\bf r}_i} V_0({\bf r} - {\bf r}_i).
\end{gather}

Here ${\boldsymbol \sigma} = \{\sigma_x,\sigma_y\}$ is the set of Pauli matrices, $v_0\approx 10^6$~m/s is the characteristic velocity of quasi-particles in graphene, $\hat{\bf p}$ is the two-dimensional momentum operator, and $V_0({\bf r - r}_i)$ is the potential of a single scattering center located at ${\bf r}_i$. The eigenstates of $\hat H_0$ represent quasiparticles belonging to the conduction ($\lambda = +1 $) and valence ($\lambda = -1 $) bands, $|{\bf p} \lambda \rangle = e^{i {\bf pr}/\hbar} \{1/\sqrt{2}, \lambda e^{i \theta_{\bf p}}/\sqrt{2} \}$, where $\theta_{\bf p}$ is an angle between momentum $\bf p$ and $x$-axis.

The electron-field interaction causes the direct interband transitions allowed in the first-order perturbation theory. The corresponding real part of the interband conductivity, ${\mathrm Re}[\sigma_{inter}]$, can be found from, e.g., the Fermi golden rule with the following result
\begin{equation}
\label{Typical-inter}
{\mathrm Re}\left[ \sigma_{inter} \right] = \sigma_q \left[f_v(-\hbar \omega/2) - f_c (\hbar \omega/2) \right].
\end{equation}
Here $\sigma_q = e^2/4\hbar$ is the universal optical conductivity of clean graphene, $f_v(\varepsilon)$ and $f_c(\varepsilon)$ are the electron distribution functions in conduction and valence bands. Below we assume them to be quasi-equilibrium Fermi functions, $f_\lambda(\varepsilon) = \left\{ 1+\exp[(\varepsilon - \mu_\lambda)/T]\right\}^{-1}$. In equilibrium, $\mu_c = \mu_v = \varepsilon_F \geq 0$, while for symmetrical pumping $\mu_c = -\mu_v = \varepsilon_F$, where $\varepsilon_F$ is the (quasi) Fermi energy. Then, one readily notes that the conductivity of the pumped graphene due to the direct interband transitions is negative at frequencies $\hbar \omega < 2 \varepsilon_F$ 

The presence of scattering potential $\hat V_S({\bf r})$ leads to the indirect electron transitions and, thus, intraband Drude absorption. The corresponding electron transition amplitude $V_{{\bf pp}'}^{cc}$ is readily found from the second-order perturbation theory considering (\ref{Field}) and (\ref{Scattering}) as perturbations
\begin{equation}
\label{V-intra}
V_{{\bf pp}'}^{cc} = \frac{e}{2c}\frac{(\bf {v_p} - {\bf v}_{{\bf p}'}, {\bf A}_0)}{\hbar \omega}  \langle {\bf p}' c | \hat V_S | {\bf p} c \rangle.
\end{equation}
Similar expression can be written for the electron transitions in the valence band. 

However, one can also consider a second-order process, where the photon absorption/emission is followed or preceded by the {\it interband} electron scattering. Its amplitude is similarly found from the second-order perturbation theory, which results in 
\begin{equation}
\label{V-inter}
V_{{\bf pp}'}^{cv} = \frac{e}{2c}\frac{(\bf {v_p} + {\bf v}_{{\bf p}'}, {\bf A}_0)}{\hbar \omega}  \langle {\bf p}' c | \hat V_S | {\bf p} v \rangle.
\end{equation}

Applying the Fermi golden rule for these indirect inter- and intraband transitions, one can find the real part of dynamic conductivity. We mark it with the upper index '$D$' (Drude) to distinguish from the interband conductivity due to the direct transitions (\ref{Typical-inter}):
\begin{widetext}
\begin{equation}
\label{Sigma-intra}
\mathrm{Re} \left[ {\sigma^D_{\mathrm{intra}}} \right] =
\frac{e^2}{4 \hbar} \frac{2\pi g}{\hbar \omega^3} \sum\limits_{ {\bf p}, {\bf p}', \lambda}
{
\left[ f_\lambda( \varepsilon_{\bf p} ) - f_\lambda( \varepsilon_{{\bf p}'}) \right]
\delta \left[ \varepsilon_{{\bf p} \lambda} + \hbar \omega - \varepsilon_{{\bf p}'\lambda} \right]}
\left| \langle {\bf p}' \lambda | \hat V_S | {\bf p} \lambda \rangle \right|^2
\left({\bf v}_{{\bf p}'} - {\bf v}_{\bf p} \right)^2,
\end{equation}

\begin{equation}
\label{Sigma-inter}
\mathrm{Re} \left[ {\sigma^D_{\mathrm{inter}}} \right] =
\frac{e^2}{4 \hbar} \frac{2\pi g}{\hbar \omega^3}
\sum\limits_{{\bf p},{\bf p}'} {\left[ f_v ( \varepsilon_{\bf p} )- f_c ( \varepsilon_{{\bf p}'} ) \right]
\delta \left[ \hbar \omega - {\varepsilon }_{\bf p} - \varepsilon_{{\bf p}'} \right]}
{\left| \langle {\bf p}' c | \hat V_S | {\bf p} v \rangle \right|^2
\left( {\bf v}_{{\bf p}'} + {\bf v}_{\bf p} \right)^2}.
\end{equation}
\end{widetext}
Here ${\bf v_p} = v_0 {\bf p}/p$ is the quasiparticle velocity and $g=4$ is the spin-valley degeneracy factor. In the classical limit $\hbar \omega \ll \{k_B T, \varepsilon_F\}$, equation (\ref{Sigma-intra}) naturally reduces to the conductivity obtained with the Boltzmann theory. From the other hand, Eq.~ (\ref{Sigma-intra}) is valid until $\omega \gg \tau_p^{-1}$, where $\tau_p$ is the quasiparticle momentum relaxation time. We will focus on this frequency range in the following.  Equation~(\ref{Sigma-inter}) cannot be obtained from simple kinetic equation as it involves the interband transitions. This is the central equation of the present letter. Further we will analyze it and compare the magnitudes of $\mathrm{Re}\left[ \sigma^D_{\mathrm{intra}} \right]$ and $\mathrm{Re} \left[ \sigma^D_{\mathrm{inter}} \right]$. 

A key difference between the inter- and intraband Drude conductivities can be seen from the energy conservation law. For the interband energy conservation in Eq.~(\ref{Sigma-inter}) to be fulfilled, the transmitted momentum ${\bf q} = {\bf p}' - {\bf p}$ should be small, namely $q v_0 \le \hbar \omega$. In the intraband process~(\ref{Sigma-intra}), oppositely, $q v_0 \ge \hbar \omega$ is required. Therefore, the indirect interband transitions are favored by scattering potentials with large Fourier components $V_{\bf q}$ at small ${\bf q}$. For the indirect interband transitions to dominate over intraband ones, the Fourier components of the scattering potential should be either singular at $q \rightarrow 0$ or decrease abruptly as $q$ increases.

Consider the scattering by random defects with the Gaussian correlations, such that the potential energy correlator is $\langle V ({\bf r}) V({\bf r}') \rangle = \overline {V^2} \exp\left[ { - |{\bf r} - {\bf r}'|^2/l_c^2} \right]$, where $l_c$ is the correlation length and $\overline {V^2}$ is the average square of scattering potential. The squared modulus of the scattering matrix element appearing in the Fermi golden rule is calculated as~\cite{Vasko-Ryzhii,Das-Sarma-correlated}
\begin{equation}
\left| \langle {\bf p}' \lambda' | \hat V_S | {\bf p} \lambda \rangle \right|^2 = \pi l_c^2 \overline {V^2} e^{-(q l_c/2)^2} \frac{1 + \lambda \lambda' \cos\theta_{{\bf pp}'}}{2}.
\end{equation}

\begin{figure}[h]
\center{\includegraphics[width=0.95\linewidth]{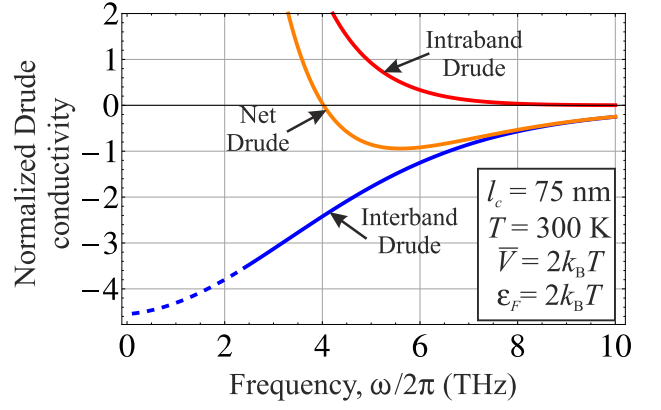}}
\caption{Calculated frequency dependencies of the interband, intraband and net Drude conductivity (normalized by $\sigma_q$) in pumped graphene with quasi-Fermi energy $\varepsilon_F = 50$ meV. The distribution of impurities is Gaussian. The dashed line indicates the region $\omega < \tau_{p = \varepsilon_F/v_0}^{-1}$, where our calculations are not rigorous.}
\label{F2}
\end{figure}

Passing to the elliptic coordinates in Eq.~(\ref{Sigma-inter}) and evaluating the integrals, one obtains the following relation for the interband Drude conductivity:
\begin{multline}
{\mathrm Re} \left[ {\frac{\sigma_{inter}^D}{\sigma_q}} \right] = \frac{\overline{V^2} l^2_c}{\hbar^2 v^2_0}\times \\
\left[f_v(-\hbar \omega/2) - f_c(\hbar \omega/2)\right] {\cal J}_1 \left(\frac{\omega l_c}{2v_0}\right),
\end{multline}
where ${\cal J}_1(x)$ is a dimensionless integral
\begin{equation}
{\cal J}_1(x) = \int_0^1{dt e^{- t x^2} \sqrt{1-t} \left(1 - \sqrt{1-t} \right)}
\end{equation}
with the following asymptotic values 
\begin{equation}
{\cal J}_1 (x) = \displaystyle{\left\{ 
\begin{array}{l l}
1/6, x \ll 1,\\
x^{-4}/2, x \gg 1.
\end{array}\right.}
\end{equation} 
Considering the intraband Drude conductivity associated with the correlated disorder, one can show that the energy-momentum restriction $q v_0 > \hbar \omega$ yields a small factor $\exp\left\{ - (\omega l_c/2v_0)^2\right\}$ in the expression for ${\mathrm Re} \left[ \sigma^D_{intra} \right]$. This results in an abrupt drop beyond $1/\omega^2$ in intraband Drude absorption with increasing frequency. There is no such a small factor in the expression for the interband Drude absorption. As a result, the net Drude conductivity  ${\mathrm Re}\left[\sigma_{intra}^D + \sigma_{inter}^D\right]$ can become negative in some frequency range below $2 \varepsilon_F/\hbar$. This is illustrated in Fig.~\ref{F2} where we plot separate contributions of inter- and intraband scattering processes to the Drude conductivity.

\begin{figure}[h]
\center{\includegraphics[width=0.95\linewidth]{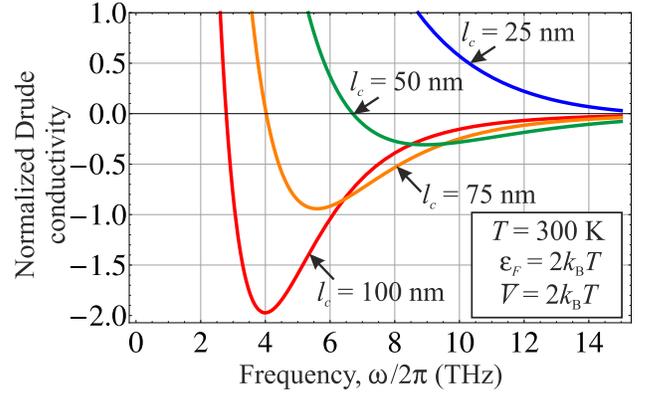}}
\caption{Real part of the net Drude conductivity (normalized by $\sigma_q$) vs. frequency at different correlation lengths $l_c$.}
\label{F3}
\end{figure}

At a constant value of  $\overline{V^2}$, the minimum in the net Drude conductivity shifts to even lower values with increasing the correlation length $l_c$. This is mainly due to abrupt drop in 'normal' Drude absorption ${\mathrm Re}[\sigma_{intra}^D] \propto e^{ - (\omega l_c/2v_0)^2}$. The minimum is also shifted toward smaller frequencies as $l_c$ increases. Roughly, the minimum is achieved at $\omega \sim v_0/l_c$. These two trends are illustrated in Fig.~\ref{F3}, where we plot the net Drude conductivity vs. frequency at different correlation lengths.

One can also find the contributions to the interband Drude conductivity associated with the scattering by random {\it uncorrelated} charged impurities with average density $n_i$. The result reads as follows
\begin{multline}
{\mathrm Re} \left[ \frac{\sigma^D_{inter}}{\sigma_q } \right] = 8 \pi \alpha_c^2 \left( \frac{v_0 \sqrt{n_i}}{\omega} \right)^2 \times \\
\left[f_v(-\hbar \omega/2) - f_c(\hbar \omega/2)\right] {\cal J}_2\left(\frac{q_S v_0}{\hbar \omega}\right),
\end{multline}
where
\begin{equation}
{\cal J}_2 (x) =\int_0^1{\frac{t dt\sqrt{1-t^2}}{(t+x)^2}\left( 1 - \sqrt{1-t^2} \right)},
\end{equation}
$q_S = 4 \alpha_c (k_B T/v_0) \ln (1+e^{\mu_c/k_B T})(1 + e^{-\mu_v/k_B T})$ is the Thomas-Fermi (screening) momentum~\cite{Das-Sarma-plasmons}, $\alpha_c = e^2/(\kappa_0 \hbar v_0)$ is the coupling constant, and $\kappa_0$ is the background dielectric constant.

A direct numerical comparison of ${\mathrm Re} \left[\sigma^D_{inter}\right]$  and ${\mathrm Re} \left[\sigma^D_{intra}\right]$ for the uncorrelated impurities shows that account of interband processes reduces the value of Drude absorption by maximum of $\sim 5$ \% in pumped graphene for the typical values $\varepsilon_F = 50$ meV and $\kappa_0=5$ at room temperature. Hence, one should not expect a sufficient renormalization of the radiation absorption due to the interband transitions for such scattering potentials. This conclusion holds valid for any scattering potential weakly depending on ${\bf q}$ (acoustic phonon scattering is another example). 
\begin{figure}[ht]
\center{\includegraphics[width=0.95\linewidth]{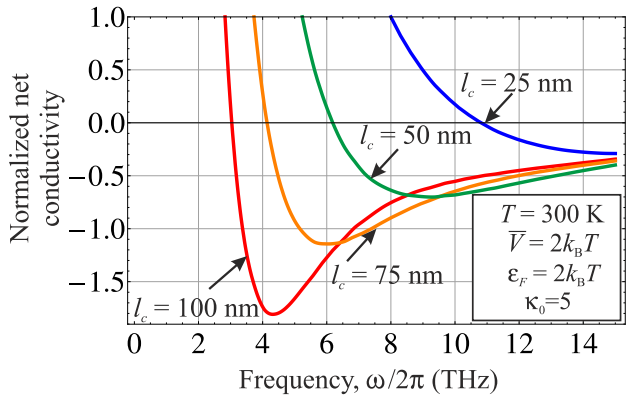}}
\caption{Real part of the normalized net conductivity of pumped graphene (including $\sigma_{inter}$, $\sigma^D_{inter}$, $\sigma^D_{intra}$, and $\sigma_{cc}$) vs. frequency for different correlation lengths $l_c$.}
\label{F4}
\end{figure}
In those cases, the phase space restriction $q < \hbar \omega/v_0$ results in the relative smallness of the interband Drude conductivity.
\begin{figure}[ht]
\center{\includegraphics[width=0.85\linewidth]{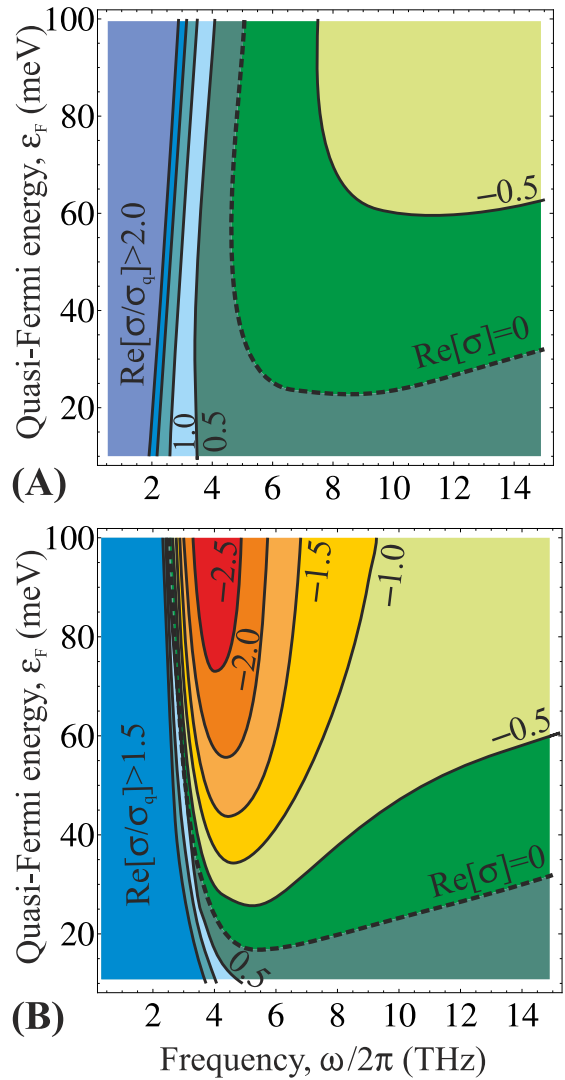}}
\caption{Color map of pumped graphene conductivity vs. frequency and quasi-Fermi energy of carriers. (A) clean graphene, conductivity is due to the direct interband transitions and the carrier-carrier scattering only. (B) graphene with the Gaussian correlated impurities. The numbers indicate the values of $\mathrm{Re}[\sigma/\sigma_q]$ on the contours. Parameters used: $T = 300$ K, $\overline V = 2 k_B T$, $l_c = 100$ nm, $\kappa_0=5$.}
\label{F5}
\end{figure}

From a practical point of view, it is important to calculate the {\it net} dynamic conductivity of pumped graphene. The optical conductivity of clean graphene incorporates the direct interband conductivity (\ref{Typical-inter}) and Drude conductivity due to carrier-phonon~\cite{Photoconductivity-graphene,Vasko-Ryzhii} and carrier-carrier scattering~\cite{CCScattering}; we denote the latter by $\sigma_{ph}$ and $\sigma_{cc}$, respectively. It was shown~\cite{CCScattering} that in the high-frequency range $\omega \gg \tau_p^{-1}$ one typically has $\sigma_{cc} \gg \sigma_{ph}$. In samples with correlated disorder, the contributions (\ref{Sigma-inter}) and (\ref{Sigma-intra}) sum up with the conductivity of clean graphene.

Figure~\ref{F4} shows the examples of the  net dynamic conductivity in such samples calculated for  different correlation lengths $l_c$. As seen from Fig.~4, even in the presence of strong carrier-carrier scattering, the interband limit $-\sigma_q$ of the negative dynamic conductivity can be surpassed due to indirect interband transitions.

In Fig.~\ref{F5}, we compare the dynamic conductivities of a clean pumped graphene layer and a graphene layer with the Gaussian correlated impurities at different frequencies and quasi-Fermi energies. The presence of the interband Drude transitions significantly broadens the domain of negative conductivity, particularly, toward the lower frequencies. The negative dynamic conductivity below $-\sigma_q$ manifests the possibility of optical amplification above 2.3\% by a single layer of pumped graphene.

In conclusion, we have studied the processes of photon emission/absorption in graphene enabled by indirect interband transitions. The scattering by the Gaussian correlated disorder results in the interband Drude conductivity exceeding the 'normal' intraband conductivity. Under conditions of population inversion, this leads to the negative net Drude conductivity and radiation amplification above 2.3\% by a single graphene sheet. The fabrication of artificial correlated disorder, e.g. by selective absorption of atoms~\cite{Correlated-transport} or nanoperforation~\cite{Nano-perforation}, opens a way for the novel light-emitting graphene-based structures. 

The work was supported by the Russian Scientific Foundation(Project \#14 29 00277) and by the Japan Society for Promotion
of Science (Grant-in-Aid for Specially Promoting Research \#23000008), Japan

\bibliography{Biblio}

\end{document}